\documentclass{eslab53}

\usepackage{color}
\usepackage{xspace}
\usepackage{hyperref}
\usepackage[utf8]{inputenc}
\usepackage[normalem]{ulem}

\editors{J.H.J.~de Bruijne}
\publisher{Zenodo}
\conference{ESLAB \#53: The {\it Gaia} Universe}
\conferencedate{2019}

\def\Gaia{\textit{Gaia}\xspace}
\def\gmag{\ensuremath{G}\xspace}
\def\gbp{\ensuremath{G_{\rm BP}}\xspace}
\def\grp{\ensuremath{G_{\rm RP}}\xspace}
\def\Ks{\ensuremath{K_{\rm s}}\xspace}
\def\DeltaW{\ensuremath{\Delta W_{G2M}}\xspace}

\title{Long-period variables in the \Gaia era}
\author{
        Nami Mowlavi$^{1,2}$,
        Michele Trabucchi$^{3}$,
        Thomas Lebzelter$^{4}$
}

\affiliation{$^{1}$ Department of Astronomy, University of Geneva, Ch. des Maillettes 51, 1290 Versoix, Switzerland\\
             $^{2}$ Department of Astronomy, Beijing Normal University, 19 XinJieKouWai St., HaiDian District, 100875 Beijing, China\\
             $^{3}$ Dipartimento di Fisica e Astronomia Galileo Galilei Universit\`a di Padova, Vicolo dell'Osservatorio 3, 35122 Padova, Italy\\
             $^{4}$ Department of Astrophysics, University of Vienna, Tuerkenschanz\-strasse 17, 1180 Vienna, Austria
             }

\shorttitle{LPVs in \Gaia DR2}
\shortauthors{Nami Mowlavi et al.}

\abs{The second \Gaia data release (DR2, spring 2018) included a unique all-sky catalogue of large-amplitude long-period variables (LPVs) containing Miras and semi-regular variables.
These stars are on the Asymptotic Giant Branch (AGB), and are characterized by high luminosity, changing surface composition, and intense mass loss, that make them of paramount importance for stellar, galactic, and extra-galactic studies.

An initial investigation of  LPVs in the Large Magellanic Cloud (LMC) from the DR2 catalog of LPVs has revealed the possibility to disentangle O-rich and C-rich stars using a combination of optical \Gaia and infrared 2MASS photometry.
The so-called \Gaia-2MASS diagram constructed to achieve this has further been shown to enable the identification of sub-groups of AGB stars among the O-rich and C-rich LPVs.

Here, we extend this initial study of the \Gaia-2MASS diagram to the Small Magellanic Cloud and the Galaxy, and use a variability amplitude proxy to identify LPVs from the full \Gaia DR2 archive.
We show that the remarkable properties found in the LMC also apply to these other stellar systems.
Interesting features, moreover, emerge as a result of the different metallicities between the three stellar environments, which we highlight in this exploratory presentation of \Gaia's potential to study stellar populations harboring LPVs.

Finally, we look ahead to the future, and highlight the power of the exploitation of \Gaia RP spectra for the identification of carbon stars using solely \Gaia data in forthcoming data releases, as revealed in an \textit{Image of the Week} published by the \Gaia consortium on the European Space Agency's web site.

These proceedings include three animated images that can be used as outreach material.
}

\begin{document}

\maketitle

%--------------------------------------------------------------
\section{Introduction}

%\noindent Proposal 0  (used in the initial version of these proceedings): $\Delta W\!(\mathrm{\Gaia,2MASS})$\\
%Proposal 1: $\Delta W_{BP,RP,J,K_s}$ \\
%Proposal 1: $W_\mathrm{BP,RP} - W_\mathrm{J,K_s}$ \\
%Proposal 2: $\Delta W_{Gaia-2mass}$ \\
%Proposal 3: $\Delta W_{G2M}$ 

The asymptotic giant branch (AGB) stage is the final evolutionary phase of low- and intermediate-mass stars before becoming white dwarfs.
During this phase of high luminosity (they can become brighter than Cepheids), the surface composition is chemically altered by deep convective mixing, and the star loses mass at a high rate (up to $10^{-4}$~M$_\odot$/yr).
Due to these, AGB stars play a key role in stellar and galactic evolution. 
Furthermore, their light curves show variability on time scales from tens to above a thousand of days, making them classified as long-period variables (LPVs).
This, combined with their intrinsic red colors, allows to easily identify them in large-scale surveys.
This is specifically true for the \Gaia all-sky survey, which provides, among other data, \gmag photometry in a main broad optical band, and \gbp and \grp photometry in blue and red bands, respectively.

In \Gaia Data Release 2 (DR2), \citet{Mowlavi_etal18} presented a catalogue of 151'761 LPV candidates with $G$ band variability amplitudes larger than 0.2~mag (amplitudes measured between the 5 and 95\% quantiles).
Despite the use of selection criteria that aimed at prioritizing low contamination over high completeness -- the completeness was estimated to be less than 50\% over the entire sky --, this DR2 catalog already doubled the number of known such large-amplitude LPVs at the time of publication.
It thereby provided a unique unprecedented catalog for the study of populations of LPVs.

Among the first applications of this catalog, our group investigated the potential of combining the optical \Gaia \gbp and \grp photometry with infrared $J$ and \Ks photometry from 2MASS.
\citet{Lebzelter_etal18} demonstrated on the DR2 LPV candidates of the Large Magellanic Cloud (LMC) that a combination of visual and infrared Wesenheit functions, hereafter called the \Gaia-2MASS Wesenheit index and denoted \DeltaW, allows to nicely identify subgroups of AGB stars according to their mass and chemistry.
We recall the definition of this index:
\begin{equation}
  \DeltaW = W_\mathrm{RP,BP-RP} - W_\mathrm{\Ks,J-\Ks}\;,
\end{equation}
\vskip -1.5mm
\noindent with
\vskip -2mm
\begin{equation}
    W_\mathrm{RP,BP-RP} = \grp - 1.3 \left( \gbp - \grp \right)
\end{equation}
\vskip -4mm
\begin{equation}
    W_\mathrm{\Ks,J-\Ks} =  K_s - 0.686 \left( J - \Ks \right)
\end{equation}
In particular, C-rich stars can be distinguished from O-rich stars in the so-called \Gaia-2MASS diagram, which plots the absolute magnitude (we used \Ks as it is less sensitive to both interstellar extinction and intrinsic LPV variability than \gmag) versus \DeltaW.
Furthermore, low-mass, intermediate-mass and massive O-rich AGB stars, as well as supergiants, populate distinct regions in this diagram, enabling to identify them in a population of LPVs with known distances.
This provides new opportunities for stellar populations studies (e.g. age, chemical composition) from the study of sub-populations of LPVs harbored therein.
The reader is referred to \citet{Lebzelter_etal18} for a presentation of the properties and potential usage of this new \Gaia-2MASS diagram. 

This ESLAB \#53 conference aimed at showing the unique advances made possible by \Gaia DR2 in various fields of astrophysics.
In this spirit, we show in these proceedings the potential of \Gaia to study populations of LPVs using the approach introduced in \citet{Lebzelter_etal18}, but extended to the entire DR2 catalog (the study in \citet{Lebzelter_etal18} was based on the restricted list of candidates published in the DR2 catalog of LPVs) and to various stellar environments (the study in \citet{Lebzelter_etal18} was limited to the LMC).
To achieve this, we first need to be able to identify LPVs in the entire DR2 archive.
This is addressed in Sect.~\ref{Sect:sample}, and applied to the LMC in Sect.~\ref{Sect:LMC}.
Populations of LPVs in the Small Magellanic Cloud (SMC) and in the Galaxy are then explored using this technique in Sects.~\ref{Sect:SMC} and \ref{Sect:Galaxy}, respectively.

\begin{figure}[t]
	\centering
	\includegraphics[width=\linewidth]{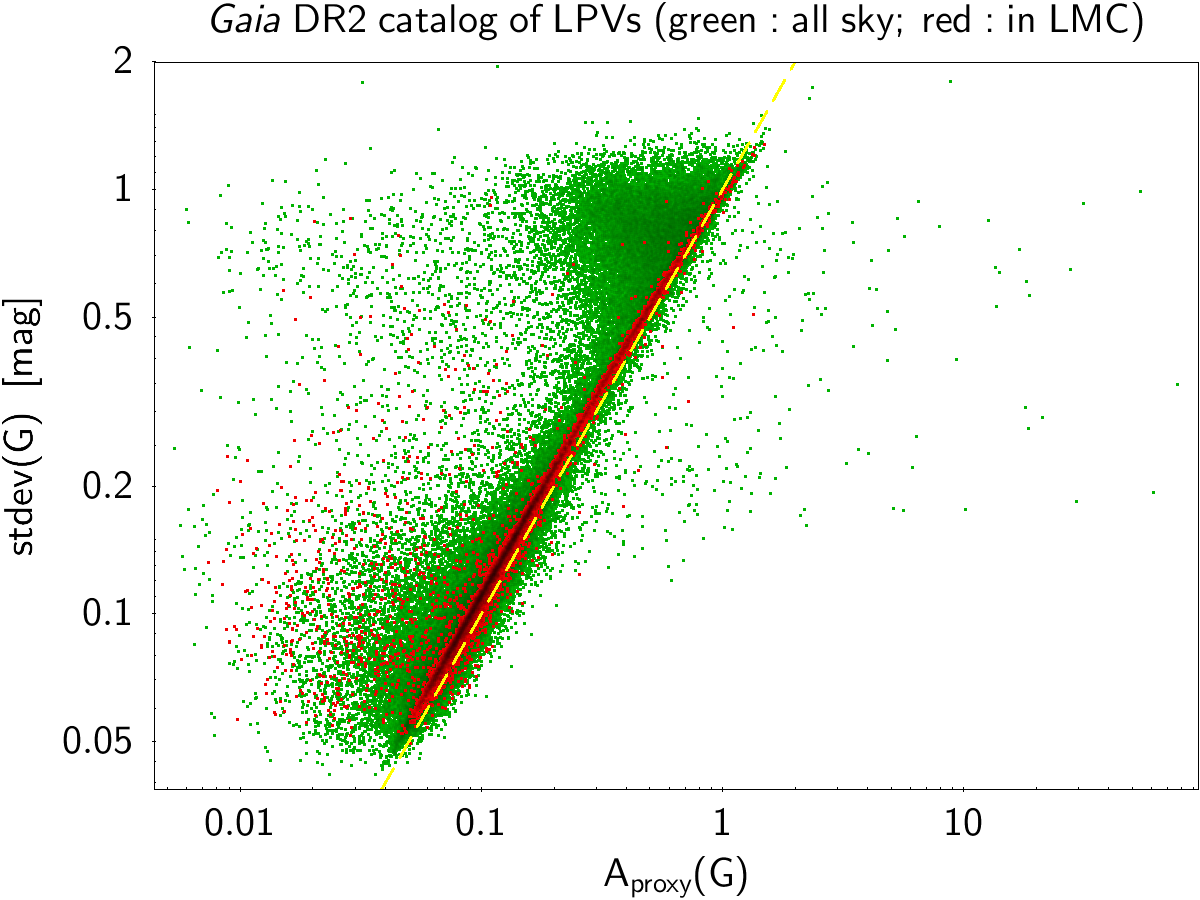}
	\caption{Variability amplitude proxy ($x$ abscissa) versus standard deviation of the \gmag magnitude time series ($y$ ordinate) for all \Gaia DR2 LPV candidates (in green).
	         The subset of candidates belonging to the LMC are shown in red.
	         The yellow dashed line is a $y\!=\!x$ diagonal line to guide the eyes. 
	}
	\label{Fig:varProxyVsStdevG}
\end{figure}

Given the paramount importance of the question of C-rich star identification for the chemical return of AGB stars and for the chemical evolution of the Galaxy (and of galaxies in general), we devote Sect.~\ref{Sect:Cstars} to a brief presentation of the unique potential of \Gaia in this respect.
We already showed in \citet{Lebzelter_etal18} the possibility of this identification using the \Gaia-2MASS diagram, as recalled above.
Its application to the SMC and the Galaxy, in addition to the LMC, is discussed in Sect.~\ref{Sect:Cstars_diagram}.
Yet, the potential of \Gaia in its future data releases is expected to be even greater, with the provision of information extracted from the epoch RP spectra of the \Gaia red spectro-photometer.
This was shown in the \Gaia \textit{Image of the Week} (IoW) of 15/11/2018 published on the \Gaia web pages of the European Space Agency (ESA)\footnote{\scriptsize \url{https://www.cosmos.esa.int/web/gaia/iow\_20181115}}, and is recalled in Sect.~\ref{Sect:Cstar_IoW}.
Section~\ref{Sect:conclusions} finally ends these proceedings with some conclusions.

For illustration and outreach purposes, movies/animated images of Fig.~\ref{Fig:G2M} (\Gaia-2MASS diagram), of Fig.~\ref{Fig:TAqr} (photometric and spectroscopic variability of an O-rich AGB star), and of Fig.~\ref{Fig:RUVir} (idem, but of a C-rich AGB star) are made available at {\footnotesize \url{https://zenodo.org/record/3237087}}.

%--------------------------------------------------------------
\section{The sample of LPV candidates}
\label{Sect:sample}

Similarly to the sample selection done in the \Gaia DR2 catalog of LPVs, we aim here at identifying \Gaia red giants with variability amplitudes larger than 0.2~mag in \gmag.
The variability amplitude proxy used to achieve this is described in Sect.~\ref{Sect:varProxy}, and the procedure to select large-amplitude LPV candidates from the full \Gaia archive is presented in Sect.~\ref{Sect:data}.

%--------------------------------------------------------------
\subsection{Variability amplitude proxy}
\label{Sect:varProxy}

\begin{figure}[t]
	\centering
	\includegraphics[width=\linewidth]{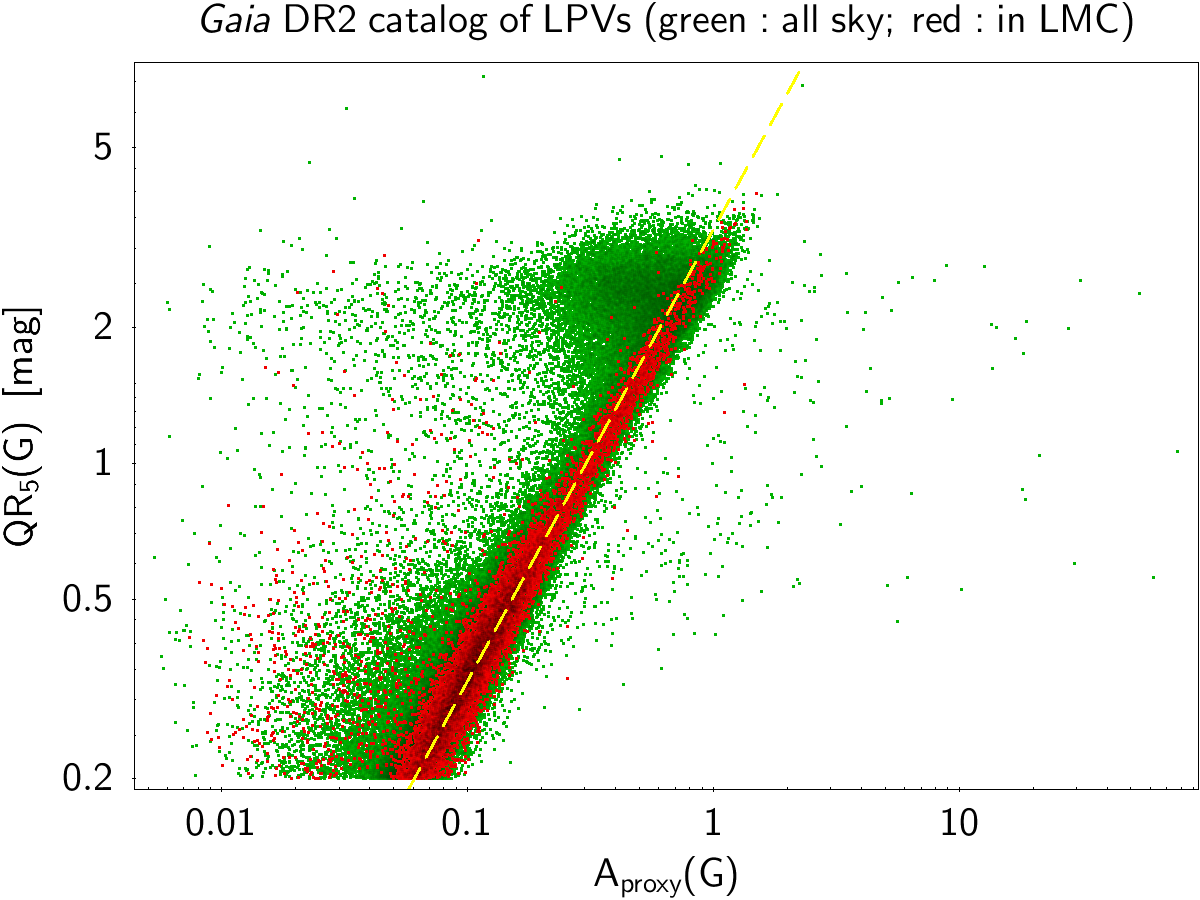}
	\caption{Same as Fig.~\ref{Fig:varProxyVsStdevG}, but with the variability amplitude proxy plotted versus the 5-95\% quantile range of the \gmag magnitude time series.
	         The yellow dashed line is $y \!=\! 3.3 *\! x$.
	}
	\label{Fig:varProxyVsTrimmedRange}
\end{figure}

In the DR2 catalog of LPVs \citep{Mowlavi_etal18}, the variability amplitude is provided through either the standard deviation or the 5-95\% quantile range $QR_5(G)$ of their \gmag magnitude time series.
These quantities are not available in the \Gaia DR2 archive for sources that have no variability-specific information in the archive.
They cannot be computed either, since the light curves of these sources are not published in DR2.

We can, however, use the uncertainty $\varepsilon(\bar{F}(G))$ of the mean \gmag flux $\bar{F}(G)$ published for each source, which contains information on both the uncertainties of the individual measurements due to noise and on the intrinsic scatter of the flux time series due to stellar variability \citep[see Sect.~5.3.5 of the \Gaia DR2 documentation in][]{BussoCacciariCarrasco_etal18}.
Since the intrinsic scatter of large-amplitude LPVs is much larger ($\ge 0.2$~mag) than the typical level of photometric uncertainties at the magnitudes considered for LPVs (few milli-magnitude uncertainties at \gmag $\lesssim$ 16~mag), the noise of the measurements has a negligible contribution to $\varepsilon(\bar{F}(G))$ for these stars.
The quantity $A_\mathrm{proxy}(G)$ defined by
\begin{equation}
  A_\mathrm{proxy}(G) = \sqrt{N_\mathrm{obs}} \; \varepsilon(\bar{F}(G)) / \bar{F}(G) \;,
\label{Eq:Aproxy}
\end{equation}
where $N_\mathrm{obs}$ is the number of (per-CCD) observations, is then a useful proxy  for the variability amplitude of stellar origin for large-amplitude LPVs.
All quantities in Eq.~\ref{Eq:Aproxy} are available in DR2.
This approach was already used by \citet{DeasonBelokurovErkal_etal17} to identify Mira variables in \Gaia DR1.

In these proceedings, we will call the variability amplitude proxy given by Eq.~\ref{Eq:Aproxy} simply as the `variability proxy'.

The relation between $A_\mathrm{proxy}(G)$ and the standard deviation of the \gmag magnitude time series of large-amplitude LPVs can be checked on the sources published in the DR2 catalog of LPVs, for which the standard deviations are available.
The result is shown in Fig.~\ref{Fig:varProxyVsStdevG}.
An almost one-to-one relation is obtained, as expected, except for a subset of sources which have a value of their (flux-based) variability proxy smaller than their (magnitude-based) standard deviation (only very few sources have their variability proxy larger than their magnitude-based standard deviation).

The relation between the variability proxy and the 5-95\% quantile range $QR_5(G)$ used in DR2 to characterize the variability amplitudes of LPVs is shown in Fig.~\ref{Fig:varProxyVsTrimmedRange}.
The majority of sources are close to the linear relation $QR_5(G) \!=\! 3.3 \!*\! A_\mathrm{proxy}(G)$.
Sources that deviate from the linear relation have a far-from-optimal sampling of measurements.
For the majority of these sources, the standard deviation (and hence the variability proxy) underestimates the variability range.
For example, for sources in the LMC shown in red in Fig.~\ref{Fig:varProxyVsTrimmedRange}, the standard deviation of sources close to the South Ecliptic Pole will have too small standard deviations due to the large number of measurements acquired during the first month of the mission\footnote{A specific scanning law, the Ecliptic Pole Scanning Law (EPSL), was used during the one-month \Gaia commissioning phase, whereby sources at the ecliptic poles were observed every six hours (spacecraft rotation period) in each of the two \Gaia fields of view.} compared to the standard deviations they would have if they had a regular sampling over their (much longer than one month) pulsation cycles.

Typical values of the variability proxy expected for LPVs can be derived from Fig.~\ref{Fig:varProxyVsTrimmedRange}.
Miras, which pulsate in the fundamental mode, were shown in \citet{Mowlavi_etal18} to have $QR_5(G) \!\gtrsim\! 1$~mag.
This translates to $A_\mathrm{proxy}(G) \!\gtrsim\! 0.3$.
Long-period variables with $QR_5(G) \!\lesssim\! 1$~mag, on the other hand, mainly consist of semi-regular variables (SRVs), pulsating in overtone modes.
They will have $A_\mathrm{proxy}(G) \!\lesssim\! 0.3$.

\begin{figure}[t]
	\centering
	\vskip 3.5mm
	\includegraphics[width=\linewidth]{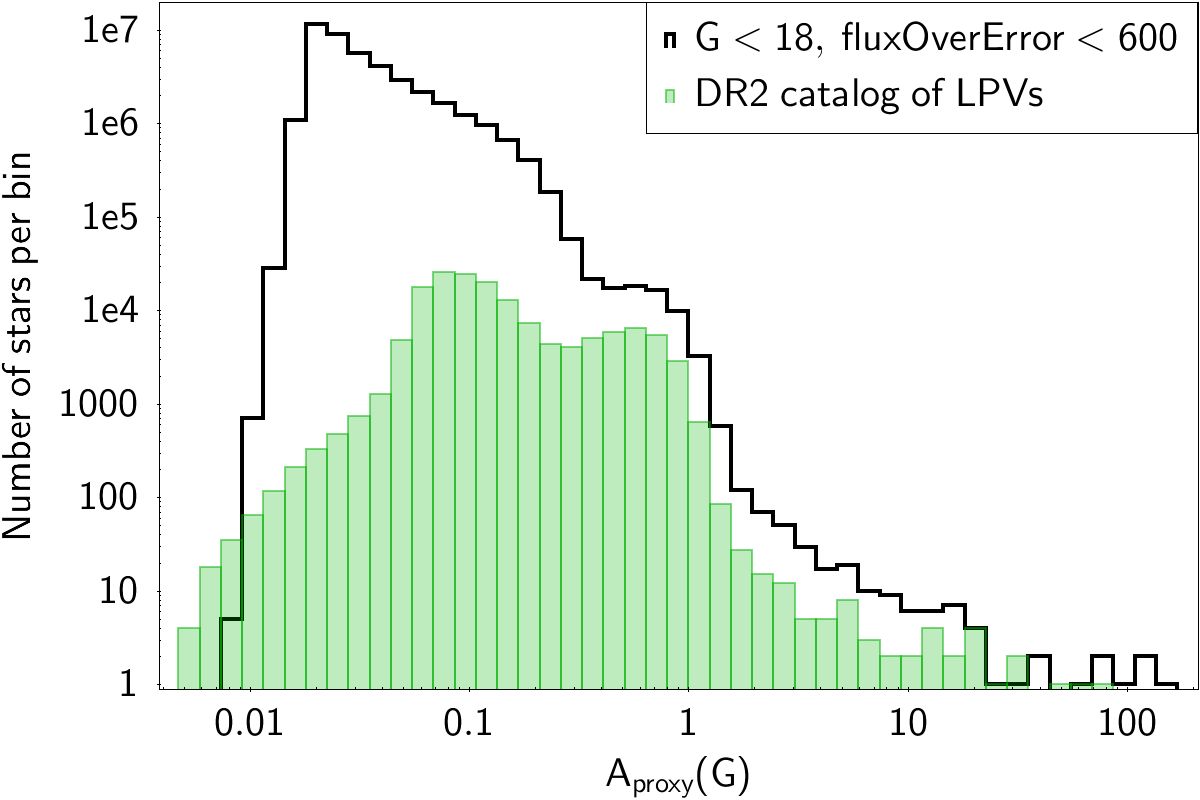}
	\caption{Black histogram: Distribution of the variability amplitude proxy of all sources in the \Gaia DR2 archive that are brighter than $G\!=\!18$~mag and that have mean \gmag flux over error ratios $\bar{F}(G)/\varepsilon(\bar{F}(G))\!<\!600$.
	         Green histogram: Same as the black histogram, but for the sources published in the specific DR2 catalog of LPVs.
	}
	\label{Fig:histo_varProxyG}
\end{figure}

\begin{table}[b]
    \centering
    \begin{tabular}{l|r|r}
    \hline
        criteria & LMC sample & SMC sample \\
        \hline
        RA & 50$^{\circ}$ to 105$^{\circ}$ & 0$^{\circ}$ to 35$^{\circ}$\\
        DEC & $-$77$^{\circ}$ to $-$61$^{\circ}$ & $-$80$^{\circ}$ to $-$65$^{\circ}$\\
        PM(RA) & 1.2 to 2.5 mas/yr & $-$0.3 to 1.6 mas/yr\\
        PM(DEC) & $-$0.8 to 1.5 mas/yr & $-$1.9 to $-$0.7 mas/yr\\
        $\varpi$ & < 0.5~arcsec & < 0.5~arcsec\\
    \hline
    \end{tabular}
    \caption{Selection criteria for stars in the LMC and SMC.}
    \label{tab:sky_selection}
\end{table}

\begin{figure}[t]
	\centering
	\includegraphics[width=\linewidth]{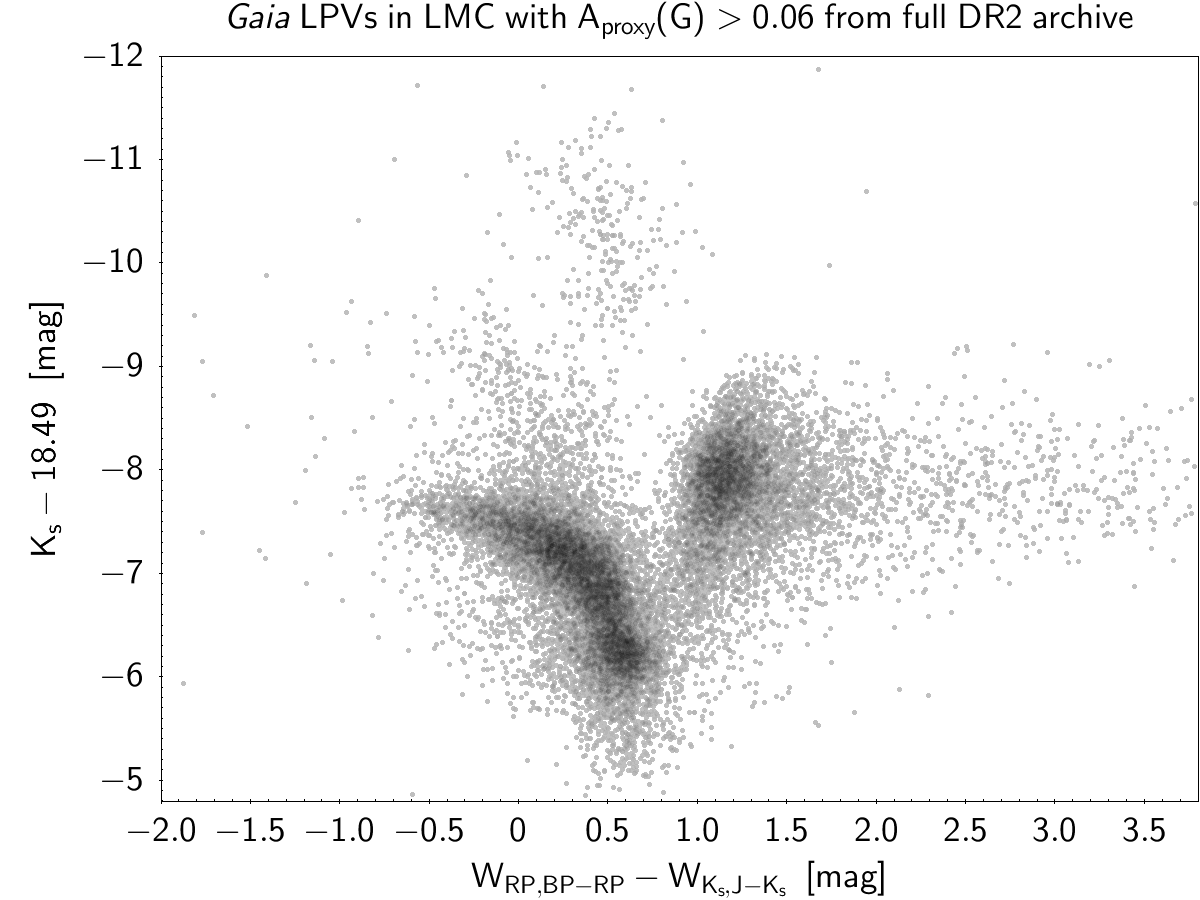}
	\caption{\Gaia-2MASS diagram of large-amplitude LPV candidates in the LMC selected from the full \Gaia DR2 archive (see text) with the condition $A_{proxy}\!>\!0.06$ for the variability proxy and having 2MASS cross matches with $J\!<\!14.5$~mag and $\Ks\!<\!14.5$~mag.
	}
	\label{Fig:LMC_Gaia2MASS_archive}
\end{figure}

%--------------------------------------------------------------
\subsection{The data}
\label{Sect:data}

The variability proxy is not an indexed column in the \Gaia archive, and thus cannot be used to select samples based on $A_\mathrm{proxy}(G)$.
We therefore  used the ratio $\bar{F}(G)/\varepsilon(\bar{F}(G))$, which is indexed in the archive, and downloaded all sources with $\bar{F}(G)/\varepsilon(\bar{F}(G))\!<\!600$ ($\rightarrow$~41'508'986 sources).
A comparison of the variability proxy distribution in this sample with the variability proxy distribution in the sample of DR2 catalog of LPVs, shown in Fig.~\ref{Fig:histo_varProxyG}, reveals that the criterion allows to retrieve the great majority (98.6\%) of the sources present in the latter catalog.
We are therefore confident that the majority of sources from the full \Gaia archive with $QR_5(G)\!>\!0.2$ are present in the downloaded sample. 
%148'714 of the 151'761 sources in the DR2 catalog of LPVs () are retrieved in this way.

We then selected as large-amplitude LPVs all red sources ($\gbp\!-\!\grp \!>\! 1.5$~mag, $\rightarrow$~19'205'377 sources) that have $A_\mathrm{proxy}(G) \!>\! 0.06$~mag ($\rightarrow$~2'952'438 sources).
This limit on the variability proxy was chosen to comply (see Fig.~\ref{Fig:varProxyVsTrimmedRange}) with the amplitude condition $QR_5(G)\!>\!0.2$~mag used in the DR2 catalog of LPVs .
Finally, we restricted the sample to sources that have 2MASS crossmatches with $J\!<\!14.5$~mag and $\Ks\!<\!14.5$~mag ($\rightarrow$~1'379'652  sources).
These conditions on the 2MASS photometry are intended to provide near-infrared data with good quality to construct \Gaia-2MASS diagrams.

This final set of 1'379'652  large-amplitude LPV candidates is used in the next sections to extract samples in the LMC, the SMC and the Galaxy.

%--------------------------------------------------------------
\section{LPVs in the Large Magellanic Cloud}
\label{Sect:LMC}

\begin{figure}[t]
	\centering
	\includegraphics[width=\linewidth]{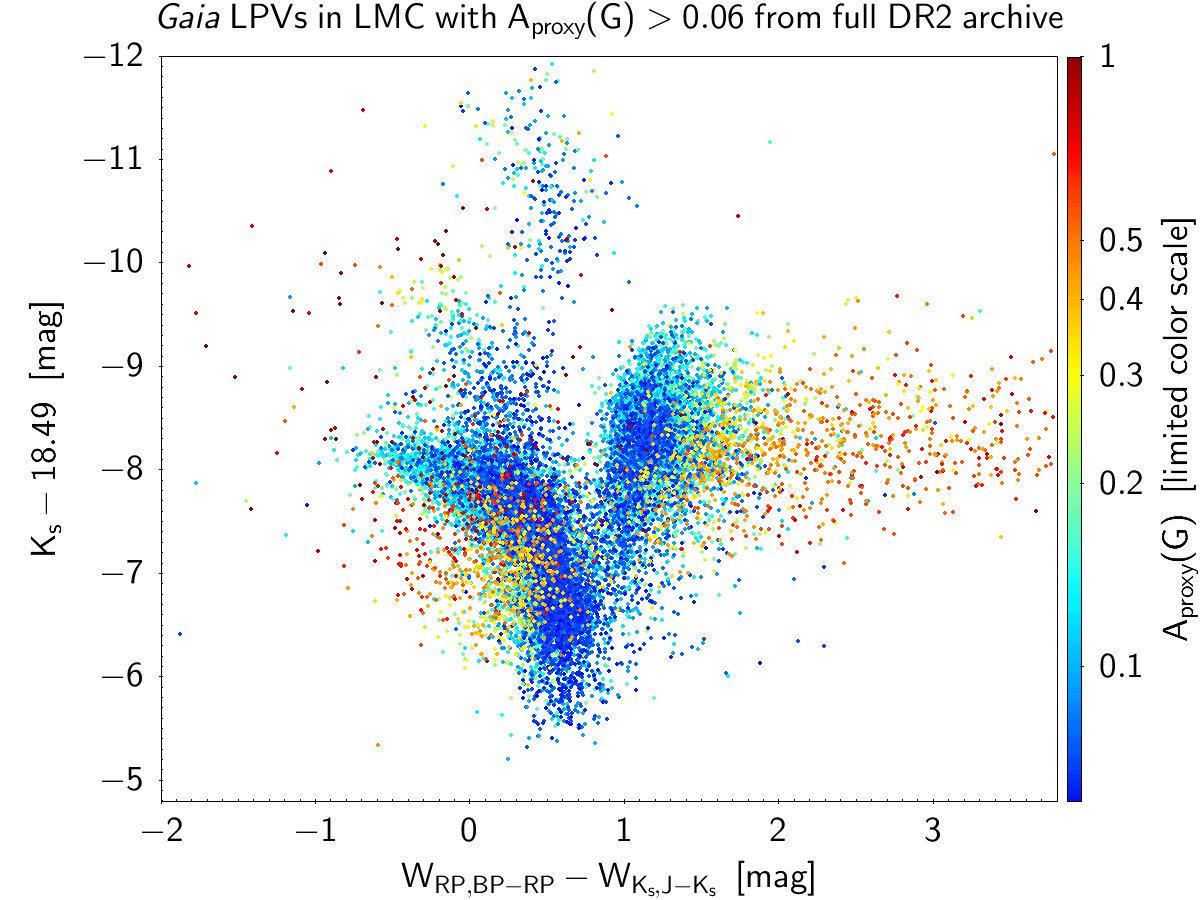}
	\caption{Same as Fig.~\ref{Fig:LMC_Gaia2MASS_archive}, but color-coded with the variability amplitude proxy of each source according to the color scale shown on the right of the figure.
	Sources with $A_\mathrm{proxy}(G)\!<\!0.06$ or $A_\mathrm{proxy}(G)\!>\!1$ are plotted with the color at the respective end of the color scale.
	The subset of sources with $A_\mathrm{proxy}(G)\!>\!0.3$ has been plotted on top of the subset with $A_\mathrm{proxy}(G)\!<\!0.3$ for visibility purposes, to ensure that the former ones are all shown on top of the latter ones.
	}
	\label{Fig:LMC_Gaia2MASS_archive_withVarProxy}
\end{figure}

\begin{figure}[ht]
	\centering
	\includegraphics[width=\linewidth]{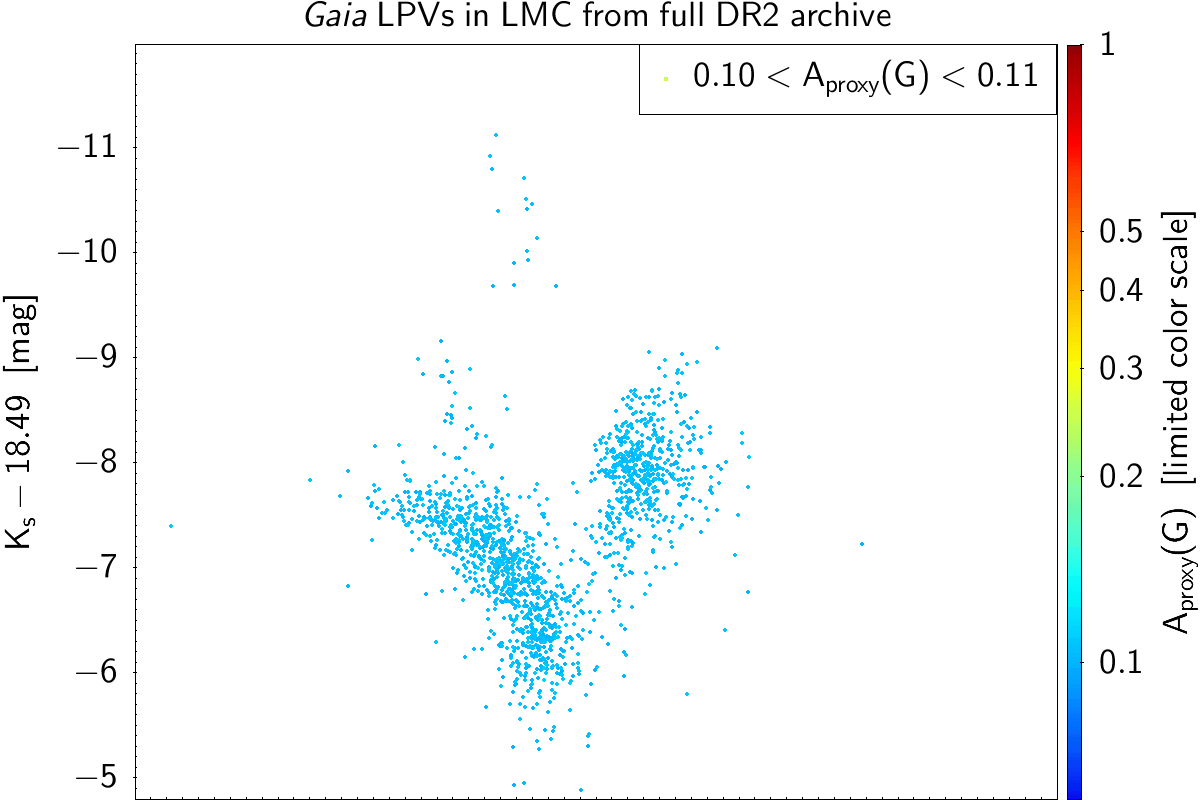}
	\vskip -0.3mm
	\includegraphics[width=\linewidth]{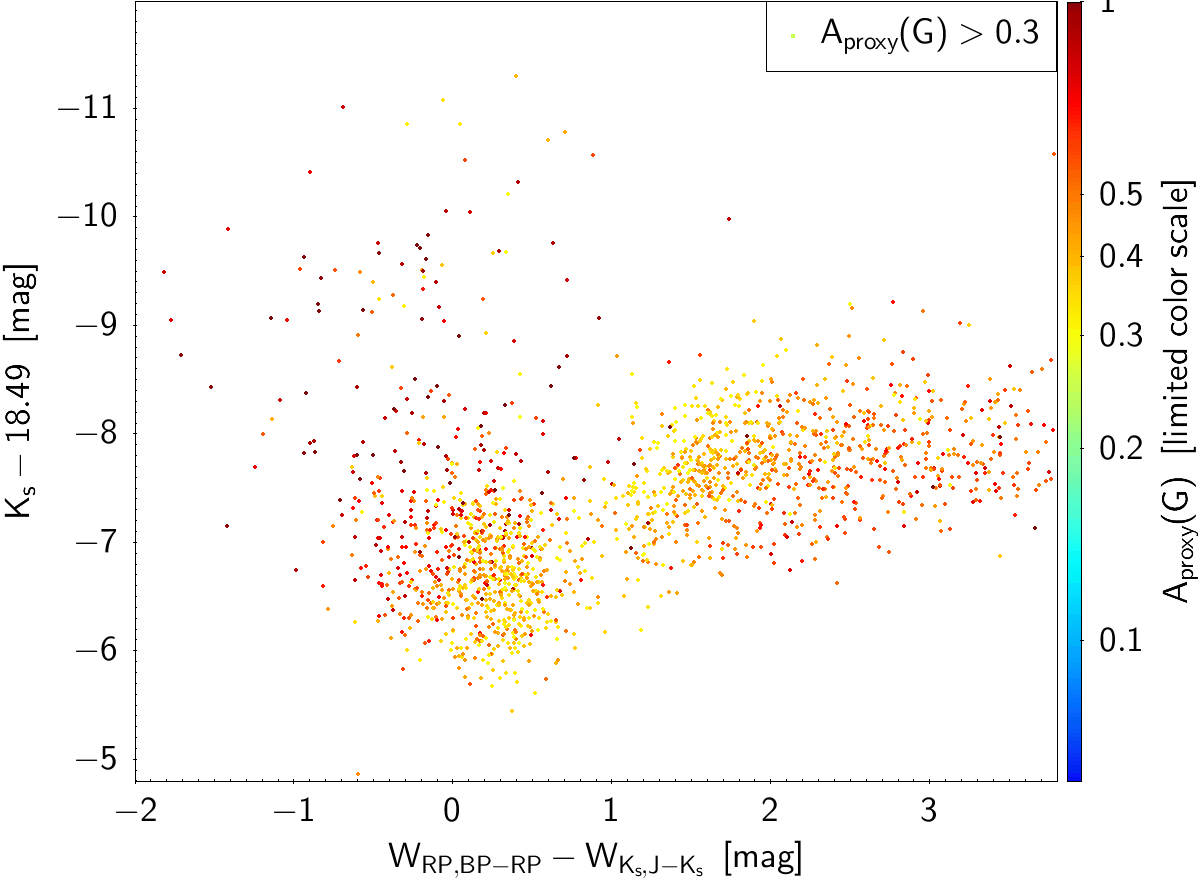}
	\caption{Same as Fig.~\ref{Fig:LMC_Gaia2MASS_archive_withVarProxy}, but for samples of red giants within restricted ranges of the variability proxy.
	The top panel is restricted to $0.10 \!<\! A_\mathrm{proxy}(G) \!<\! 0.11$ (containing mainly semi-regular variables), while the bottom panel covers $A_\mathrm{proxy}(G) \!>\! 0.3$ (containing predominantly Miras).
	}
	\label{Fig:LMC_Gaia2MASS_archive_withVarProxy_selectedVarProxies}
\end{figure}

Members of the LMC are selected on the basis of their locations in the sky, their proper motions, and their parallaxes.
The selection criteria are the same as the ones used in \citet{Lebzelter_etal18}, recalled in Table\,\ref{tab:sky_selection}.
With the conditions mentioned in Sect.~\ref{Sect:data} on \Gaia color, variability proxy and 2MASS photometry, the sample of LPVs in the LMC amounts to 20'486 sources.
This is almost twice the number of sources (10'989 sources) of the LMC published in the \Gaia DR2 catalog of LPV candidates.

The distribution in the \Gaia-2MASS diagram of these LMC LPV candidates is shown in Fig.~\ref{Fig:LMC_Gaia2MASS_archive}.
The various branches related to different stellar mass regimes identified by \citet{Lebzelter_etal18} are clearly visible for this larger sample as well. 

Figure~\ref{Fig:LMC_Gaia2MASS_archive_withVarProxy} shows the same diagram as in Fig.~\ref{Fig:LMC_Gaia2MASS_archive}, but with each point color-coded according to the value of $A_\mathrm{proxy}(G)$ of the star.
%The variables with the largest amplitudes ($A_\mathrm{proxy}(G)>0.3$) mainly consist of Mira candidates, which pulsate in the fundamental mode, while the smaller amplitude LPVs are mainly semi-regular variables (SRVs), pulsating in overtone mode(s).
%are of particular interest here, as they mainly consist of Miras which pulsate in the fundamental mode and populate sequence C in a period-luminosity diagram.
Mira candidates are plotted on top of the other stars to make them all visible.
Their distribution in the diagram seems to follow a slightly different pattern than the smaller amplitude SRVs, with Miras spreading more over the diagram than SRVs.
This is more clearly shown in Fig.~\ref{Fig:LMC_Gaia2MASS_archive_withVarProxy_selectedVarProxies}, which separates the two groups of stars, Mira candidates and SRV candidates, in the top and bottom panels, respectively.
The discussion on this feature must distinguish the case of C-rich stars (right half of the diagram at $\DeltaW \gtrsim 0.8$~mag) from the case of O-rich stars (left half of the diagram).
%This is seen in the top panel of Fig.~\ref{Fig:LMC_Gaia2MASS_archive_withVarProxy_selectedVarProxies}, which displays the distribution of stars in the \Gaia-2MASS diagram that have $0.10 < A_\mathrm{proxy}(G) < 0.11$ (which contain predominantly SRVs, see Fig.~\ref{varProxyVsTrimmedRange}), while the bottom panel displays stars with $A_\mathrm{proxy}(G) > 0.3$ (which predominantly consist of Miras). \NM{The figure disappeared, did we want to remove it (in which case this sentence must be adapted) or to keep it (in which case we have to put back the figure)? I don't remember our conclusion...}

The C-rich Miras in Fig.~\ref{Fig:LMC_Gaia2MASS_archive_withVarProxy_selectedVarProxies} (bottom panel) extend towards very large \DeltaW values compared to the SRVs (top panel).
In \citet{Lebzelter_etal18}, we named these stars `extreme C-stars', and we suggested that these are high-mass-loss objects.
The fact that only large-amplitude variables are found here confirms the widely accepted connection between large-amplitude variability and mass-loss \citep[e.g.][]{hoefner_etal18}.

Concerning O-rich Miras, many of them are found below the lowest-luminosity branch in Fig.~\ref{Fig:LMC_Gaia2MASS_archive_withVarProxy} \citep[branch (a) in][]{Lebzelter_etal18}.  
A detailed analysis of this group is beyond the scope of these proceedings.
However, we note that circumstellar reddening resulting from heavy mass loss \citep[see, e.g.,][]{mcdonald_etal18} and the appearance of significant amounts of water absorption in the near-infrared spectra \citep[see Fig.\,2 in][]{aringer_etal02} are expected to reduce the \Ks-band brightness of these stars.
Furthermore, the \Ks-band magnitudes used here are single-epoch observations, leading to a larger scatter in \Ks for large-amplitude stars.
These effects affect less SRVs because they have, on average, less circumstellar reddening and because their amplitude of variability is smaller than for Miras.
These effects also lead to a spread of the \DeltaW values of Miras, but to a lesser degree because the Wesenheit index partly corrects for the reddening effect.
In summary, the \Ks magnitudes of O-rich Miras are expected to be, on the mean, larger (fainter) than those of O-rich SRVs, and to display a larger spread due to both the larger intrinsic variability amplitudes of Miras and the single-epoch nature of the 2MASS survey.

%Furthermore, the $K$-band magnitudes used here are single-epoch observations, so that we expect to see a larger scatter for large amplitude stars.

%----------------------------------
\section{LPVs in the Small Magellanic Cloud}
\label{Sect:SMC}

\begin{figure}[t]
	\centering
	\includegraphics[width=\linewidth]{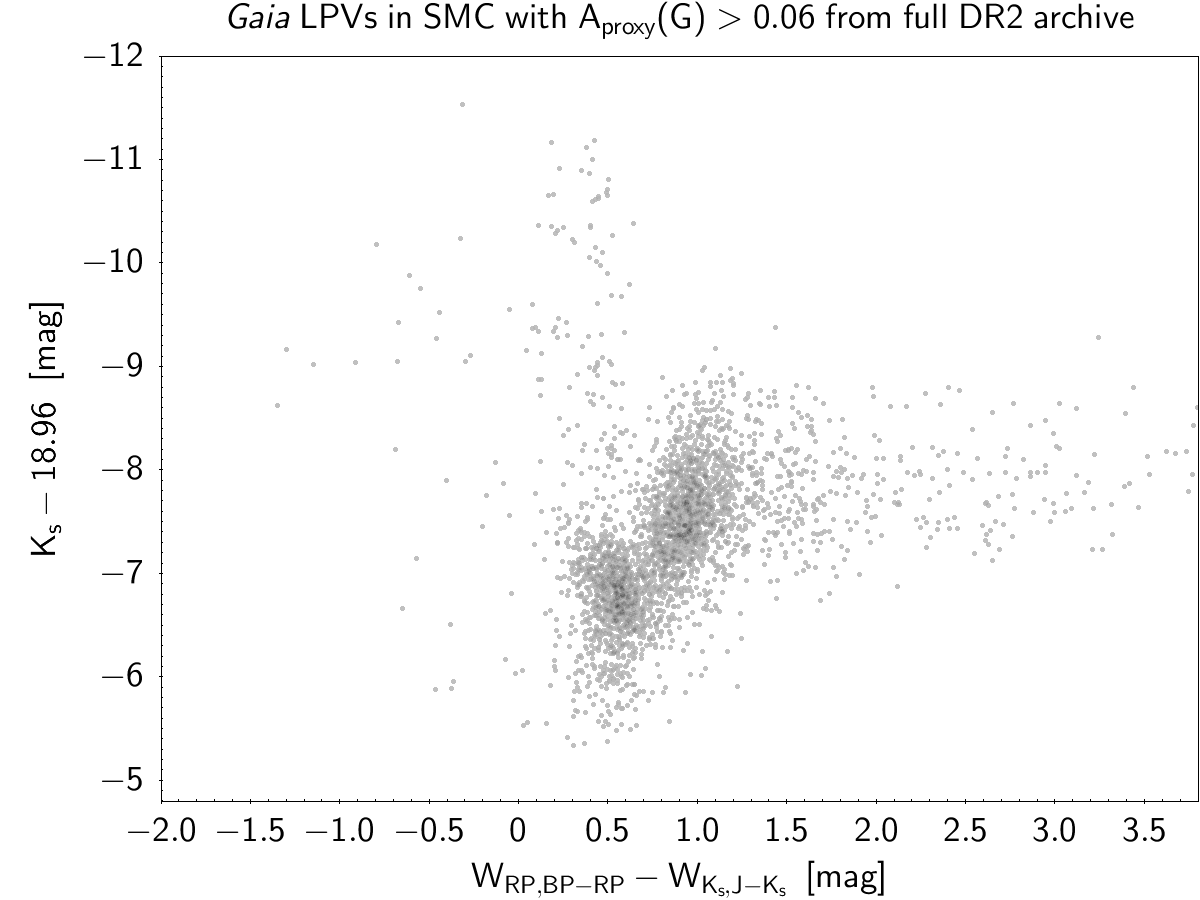}
	\caption{Same as Fig.~\ref{Fig:LMC_Gaia2MASS_archive}, but for the SMC.
	}
	\label{Fig:SMC_Gaia2MASS_archive}
\end{figure}

The selection criteria for members of the SMC are given in Table~\ref{tab:sky_selection}.
With the conditions on \Gaia color, variability proxy, and 2MASS photometry mentioned in Sect.~\ref{Sect:data} to select large-amplitude LPV candidates, we reach a sample of 3'208 sources.

\begin{figure}[t]
	\centering
	\includegraphics[width=\linewidth]{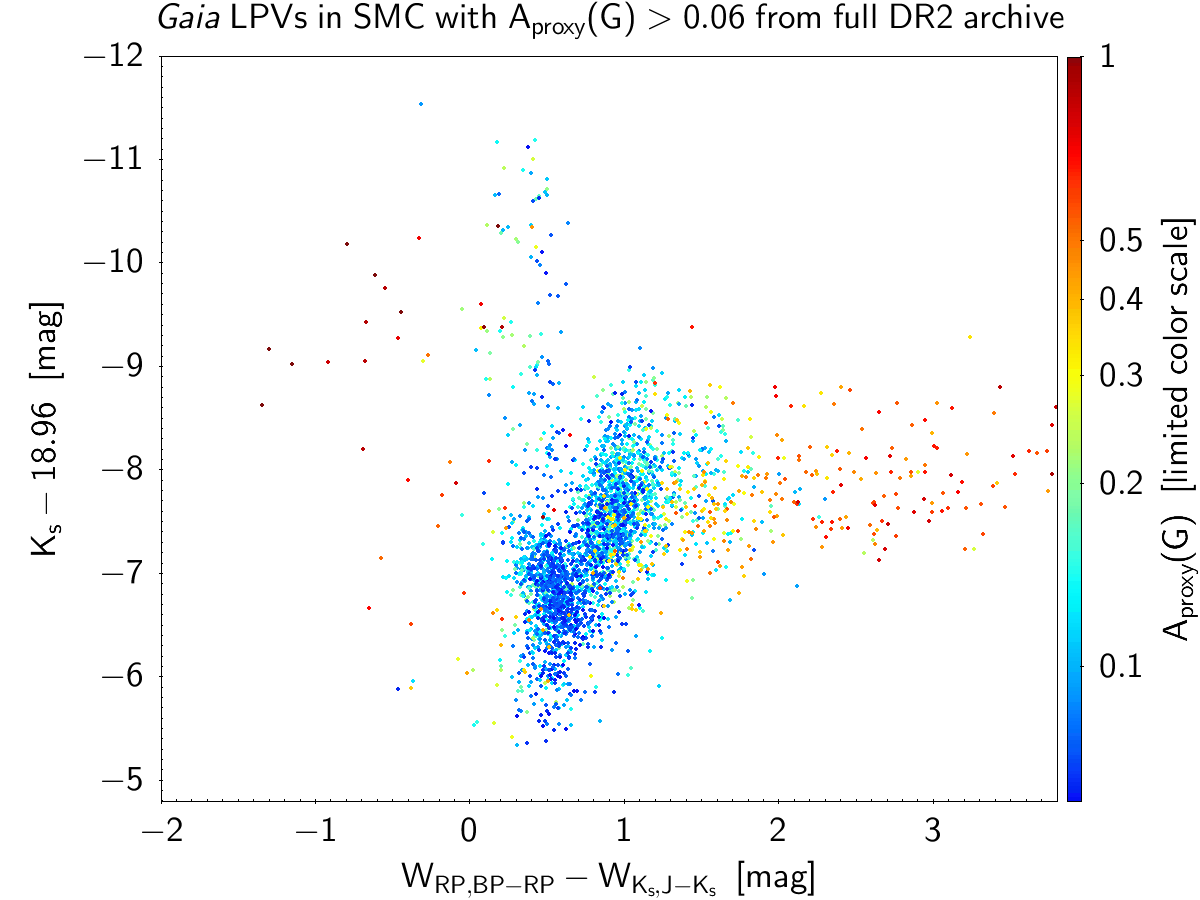}
	\caption{Same as Fig.~\ref{Fig:LMC_Gaia2MASS_archive_withVarProxy}, but for the SMC.
	}
	\label{Fig:SMC_Gaia2MASS_archive_withVarProxy}
\end{figure}

The \Gaia-2MASS diagram for the SMC is shown in Figs.~\ref{Fig:SMC_Gaia2MASS_archive} and \ref{Fig:SMC_Gaia2MASS_archive_withVarProxy}.
There are two main differences compared to the equivalent diagrams for the LMC (Figs.~\ref{Fig:LMC_Gaia2MASS_archive} and \ref{Fig:LMC_Gaia2MASS_archive_withVarProxy}, respectively).
First is the larger proportion of C-rich relative to O-rich stars in the SMC than in the LMC.
This is a well known fact, due to the lower metallicity of the SMC.
Indeed, at low metallicity, AGB stars experience more easily the third dredge-up process that mixes carbon from the deep interior where it is synthesized to the surface, and at an earlier stage on their AGB phase.
In addition, the oxygen abundance is lower in the atmosphere of an AGB star in the SMC than in the LMC, such that a smaller number of dredge-up events is needed to reach C/O>1.
As a result, the star spends an increased fraction of its lifetime as a C~star \citep[see, e.g., Fig.\,22 in][]{marigo_girardi07}.
Similar conclusions are drawn in the literature from, for example, the observation of stellar clusters in the SMC \citep[][]{frogel_etal90} and the analysis of the period-luminosity diagram of LPVs in the SMC \citep[][]{soszynski_etal11}.

The second striking feature in Fig.~\ref{Fig:SMC_Gaia2MASS_archive} is the truncation of the O-rich low-mass branch towards low \DeltaW values compared to the \Gaia-2MASS diagram of the LMC (Fig.~\ref{Fig:LMC_Gaia2MASS_archive}): there is practically no LPV with $\DeltaW \!<\! 0.2$~mag in the SMC.
This may be a direct consequence of the increased third dredge-up efficiency with decreasing metallicity, the O-rich star becoming C-rich before reaching the far left side of the \Gaia-2MASS diagram.

Let us, however, remark that the specific star formation history of the SMC, a subtle interplay between metallicity and evolution, the larger surface temperatures usually characterizing lower metallicity stars, and/or the smaller Ti abundance due to lower metallicity, all could play a role in shaping the distribution of O-rich LPVs in the \Gaia-2MASS diagram.
We further note from Fig.~\ref{Fig:SMC_Gaia2MASS_archive_withVarProxy} that almost all O-rich stars beyond the truncation limit of branch (a) on the left of the diagram, with $\DeltaW \!<\! 0.2$~mag, are Miras.

%--------------------------------------------------------------
\section{LPVs in the Galaxy}
\label{Sect:Galaxy}

The study of LPVs in the Galaxy is much more challenging than in the Magellanic Clouds.
First, because they cannot be assumed to lie at a given known distance as is the case for the LMC and the SMC.
Here \Gaia will be of key help by providing parallax measurements.
Second, because Galactic sources are affected by differential interstellar reddening that varies greatly depending on sky position and distance into the Galaxy.

We consider two samples of Galactic LPVs, selected on their relative parallax uncertainties (in addition to the LPV criteria on color, variability amplitude and 2MASS photometry precision described in Sect.~\ref{Sect:data}).
The first sample is restricted to all LPV candidates  with relative \Gaia DR2 parallax uncertainties better than 10\%, while the second sample contains a larger set with parallax uncertainties up to 15\%.
In both samples, we have to exclude potential Young Stellar Objects that can be contaminants to LPVs \citep[see][]{Mowlavi_etal18}.
We do this by imposing an upper limit on the absolute(\gmag) magnitude of LPVs, that depends on the \gbp-\grp color as defined by the dashed line in  Fig.~28 of \citet{Mowlavi_etal18}, i.e. absolute$(\gmag) \!<\! -2.6 + 1.72 * (\gbp-\grp)$.
The resulting samples contain respectively 4803 and 12'820 LPV candidates.

\begin{figure}[t]
	\centering
	\includegraphics[width=\linewidth]{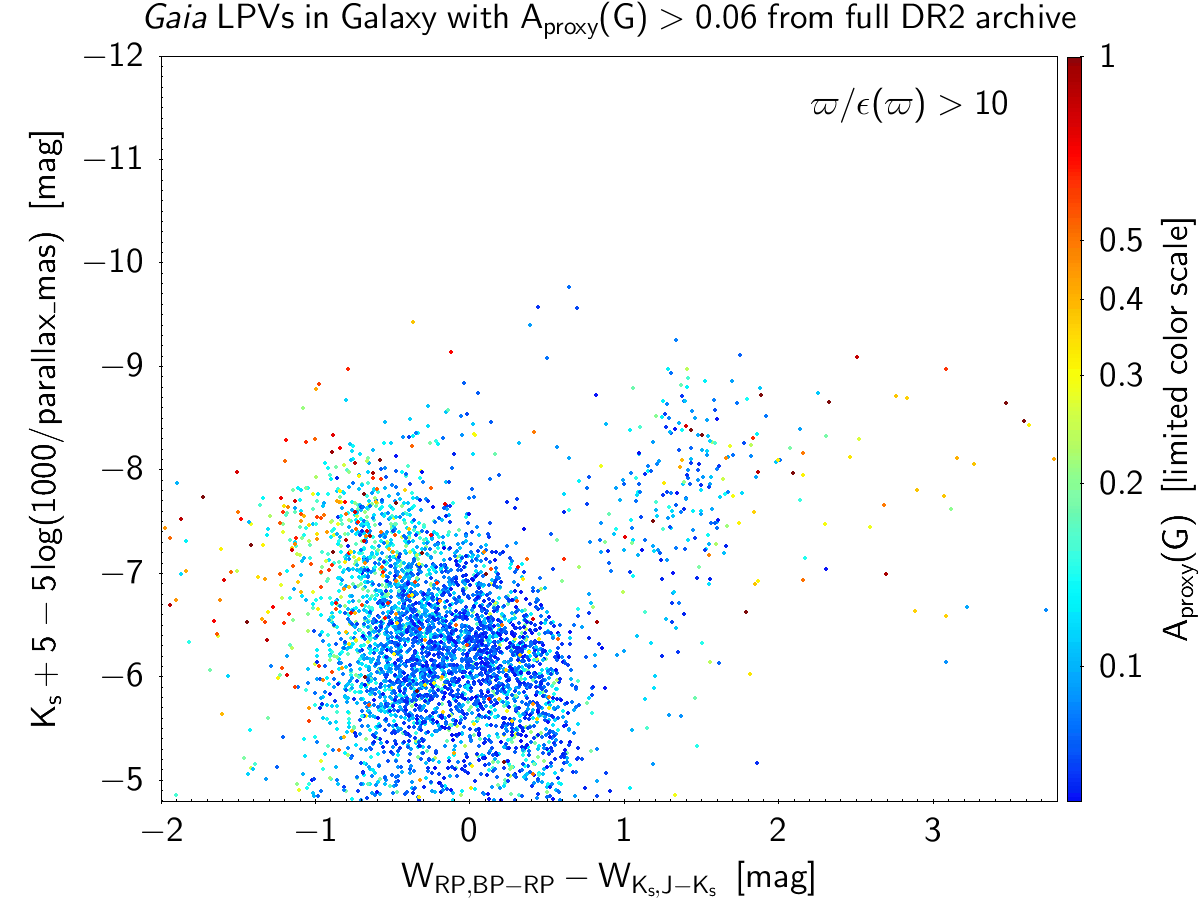}
	\caption{Same as Fig.~\ref{Fig:LMC_Gaia2MASS_archive_withVarProxy}, but for Galactic LPVs with relative DR2 parallax uncertainties better than 10\%.
	}
	\label{Fig:Gal10perc_Gaia2MASS_archive}
\end{figure}

\begin{figure}[t]
	\centering
	\includegraphics[width=\linewidth]{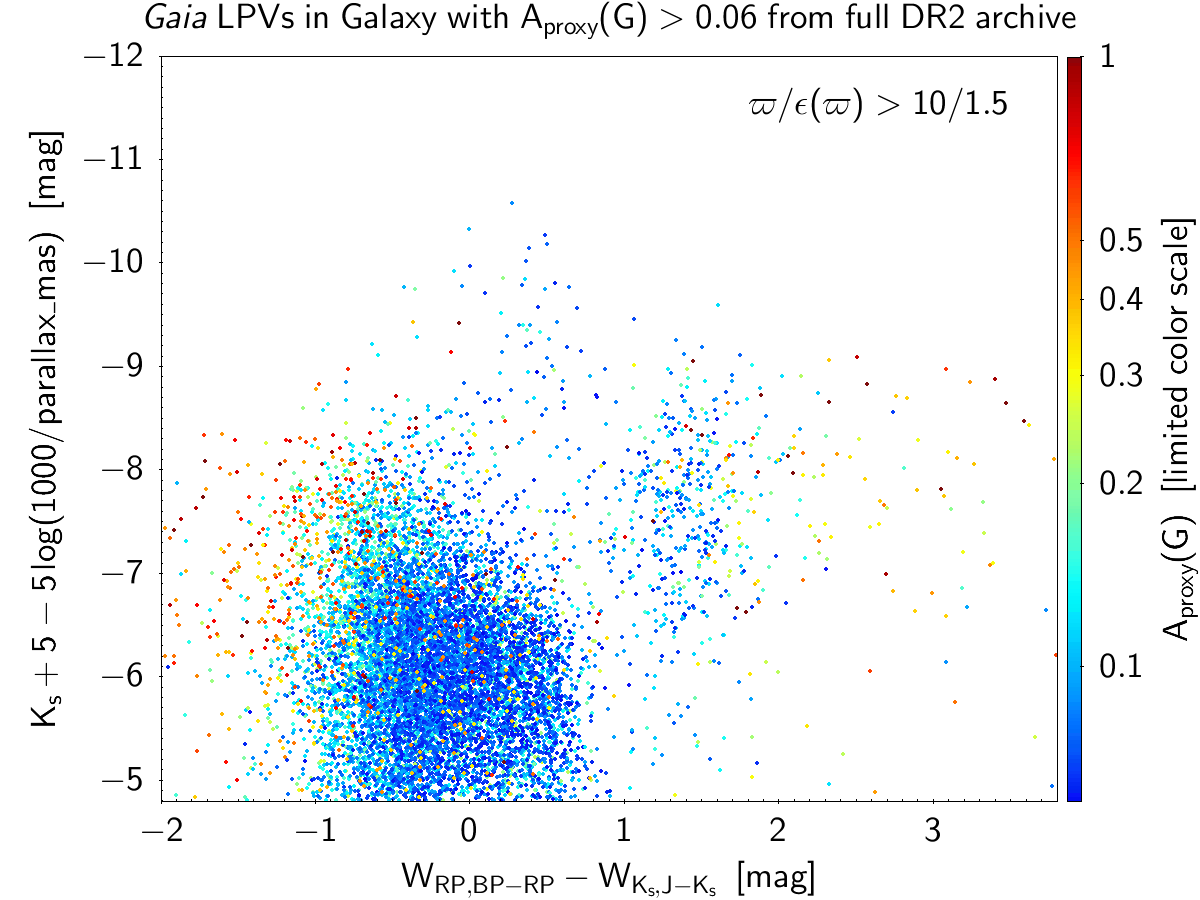}
	\caption{Same as Fig.~\ref{Fig:LMC_Gaia2MASS_archive_withVarProxy}, but for Galactic LPVs with relative DR2 parallax uncertainties better than 15\%.
	}
	\label{Fig:Gal15perc_Gaia2MASS_archive}
\end{figure}

The \Gaia-2MASS diagrams of these two samples of Galactic large-amplitude LPV candidates are shown in Figs.~\ref{Fig:Gal10perc_Gaia2MASS_archive} and \ref{Fig:Gal15perc_Gaia2MASS_archive}, respectively.
We note the following points:
\begin{itemize}
\item There are much less C-rich relative to O-rich stars in the Galaxy than in the Clouds (compare Fig.~\ref{Fig:Gal10perc_Gaia2MASS_archive} with Figs.~\ref{Fig:LMC_Gaia2MASS_archive} and \ref{Fig:SMC_Gaia2MASS_archive}).
This agrees with the decrease of third dredge-up efficiency with increasing metallicity, combined with the larger O abundances in the envelope of AGB stars, as already noticed in Sect.~\ref{Sect:SMC} when comparing the representative diagrams of the SMC and LMC.
\item The distribution of O-rich LPVs in branch (a) of the diagram (containing low-mass AGB stars as well as stars at the tip of the red giant branch) covers, at any given absolute \Ks magnitude, a much wider range of \DeltaW in the Galaxy than in the Clouds.
In Sect.~\ref{Sect:SMC}, we noticed a truncated extension of branch (a) in the SMC compared to the LMC.
Extrapolating this observed difference between the SMC and LMC to the larger metallicities (on the mean) of the Galaxy would agree with the smaller (i.e. larger negative) values of \DeltaW observed in the sample of Galactic LPVs (compare Fig.~\ref{Fig:Gal10perc_Gaia2MASS_archive} with Figs.~\ref{Fig:LMC_Gaia2MASS_archive} and \ref{Fig:SMC_Gaia2MASS_archive}).
Interstellar (and possibly circumstellar) reddening would then explain the distribution of branch (a) in Figs.~\ref{Fig:Gal10perc_Gaia2MASS_archive} and \ref{Fig:Gal15perc_Gaia2MASS_archive} towards the faint (lower) side of the diagram.
The wide distribution of branch (a) in both axes in these figures would thus result from the combined effect of O-rich AGB stars turning much later into C-rich stars in the Galaxy, if at all, and extinction (plus possibly other effects as discussed in Sect.~\ref{Sect:SMC}).
But see also the note at the end of this section on parallax determinations.
%\NM{Can we easily confirm that? Could there be other reasons?}
%\MT{This interpretation would be supported by the fact that data points seem to be pushed only leftwards, i.e., they are reddened. However, it seems to me that there is a clear trend of Aproxy increasing with decreasing WRP-WJK, which makes sense if the latter decreases as a consequence of a temperature-related shift towards red colour, rather than extinction-related reddening. I wonder how confident can we be with the absolute magnitudes here.}.
%
\item Despite the previous point, the group of C-rich stars still stands distinct from the group of O-rich stars in the \Gaia-2MASS diagram (Figs.~\ref{Fig:Gal10perc_Gaia2MASS_archive} and \ref{Fig:Gal15perc_Gaia2MASS_archive}).
This remarkable property of the \Gaia-2MASS diagram results from the fact that reddening leads to a slight displacement of the representing point in the diagram \textit{away} from the vertical line separating the two groups of stars: a reddened C-rich star will be displaced to larger \DeltaW values, while reddened O-rich stars will be displaced to smaller (negative) values of this index.
\item Only very few massive AGB and red supergiants are present in the Galactic samples of the extended Solar neighborhood considered here.
\end{itemize}

Let us finally issue a warning on the parallaxes.
Attention must be paid that systematics and biases may exist in the DR2 parallax determinations of red giants.
A bias towards the exclusion of the reddest giants in a parallax-precision-limited sample was already stressed in \cite{Mowlavi_etal18} (see in particular their Fig.~24), due to their fainter magnitudes (maximum of spectrum emission displaced to longer wavelengths in the infrared).
In addition, there were limitations in DR2 parallaxes \citep[][]{2018AA...616A...2L}.

%--------------------------------------------------------------
\section{Carbon star identification}
\label{Sect:Cstars}

The identification of C-rich and O-rich stars using \Gaia and 2MASS data is addressed in Sect.~\ref{Sect:Cstars_diagram}, with a particular focus on the effect of metallicity.
The potential of achieving the distinction in future \Gaia data releases using only \Gaia data is then highlighted in Sect.~\ref{Sect:Cstar_IoW}.

%--------------------------------------------------------------
\subsection{In the \Gaia-2MASS diagram}
\label{Sect:Cstars_diagram}

With \Gaia DR2, \citet{Lebzelter_etal18} showed how the combination of optical and infrared data allows to distinguish between O-rich and C-rich stars in the \Gaia-2MASS diagram, where these two types of LPVs occupy distinct regions.
It was shown that this optical+infrared diagram is more powerful than the infrared color-magnitude diagram usually used in the literature for this purpose.
The improvement is illustrated in Fig.~\ref{Fig:G2M}, of which an animated image is available (see caption of the figure).

\begin{figure}[t]
	\centering
	\includegraphics[width=0.95\linewidth]{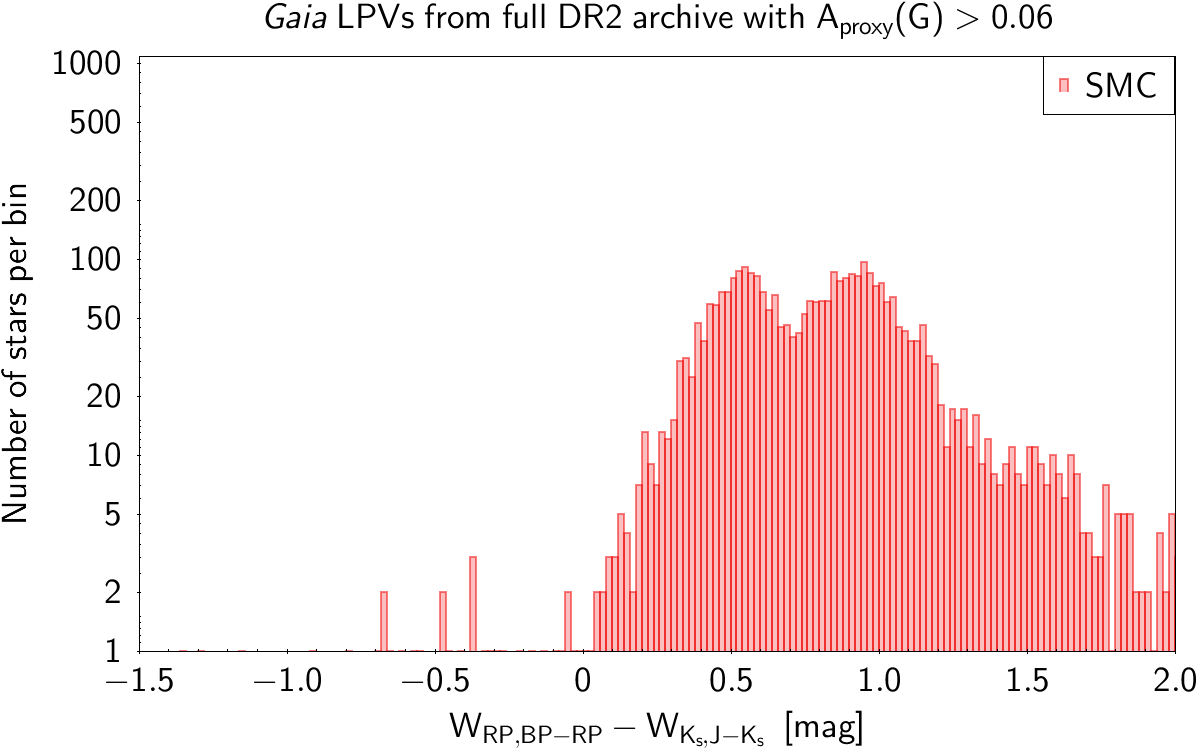}
	\vskip -6mm
	\includegraphics[width=0.95\linewidth]{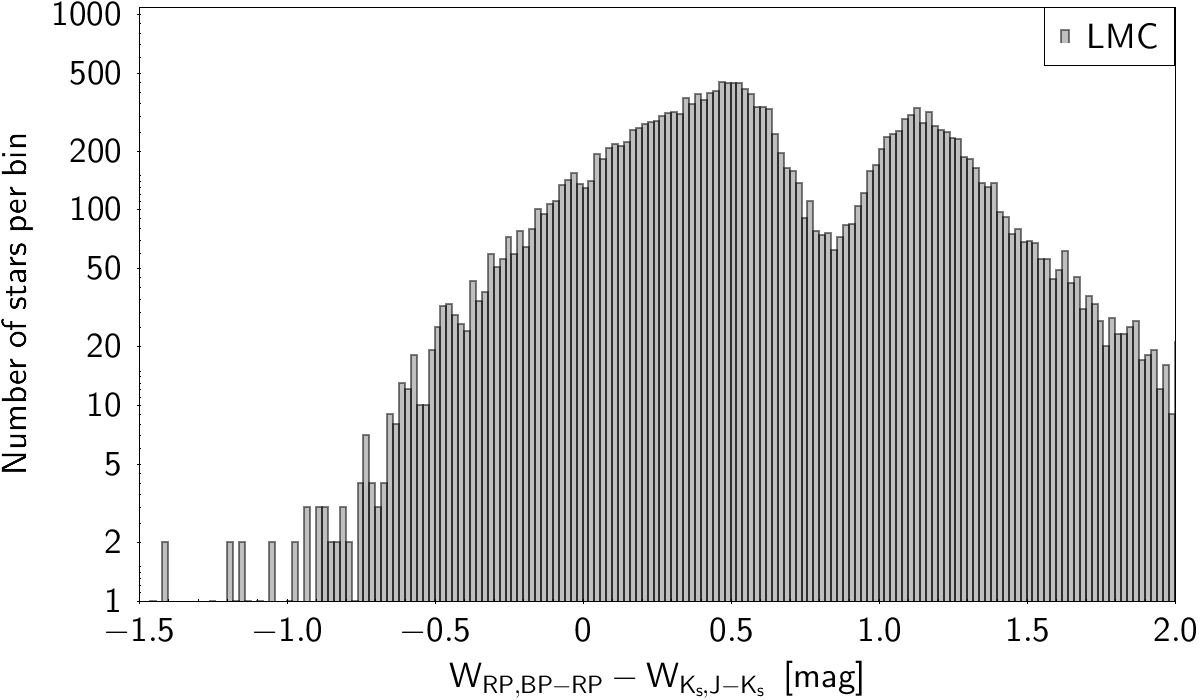}
	\vskip -6mm
	\includegraphics[width=0.95\linewidth]{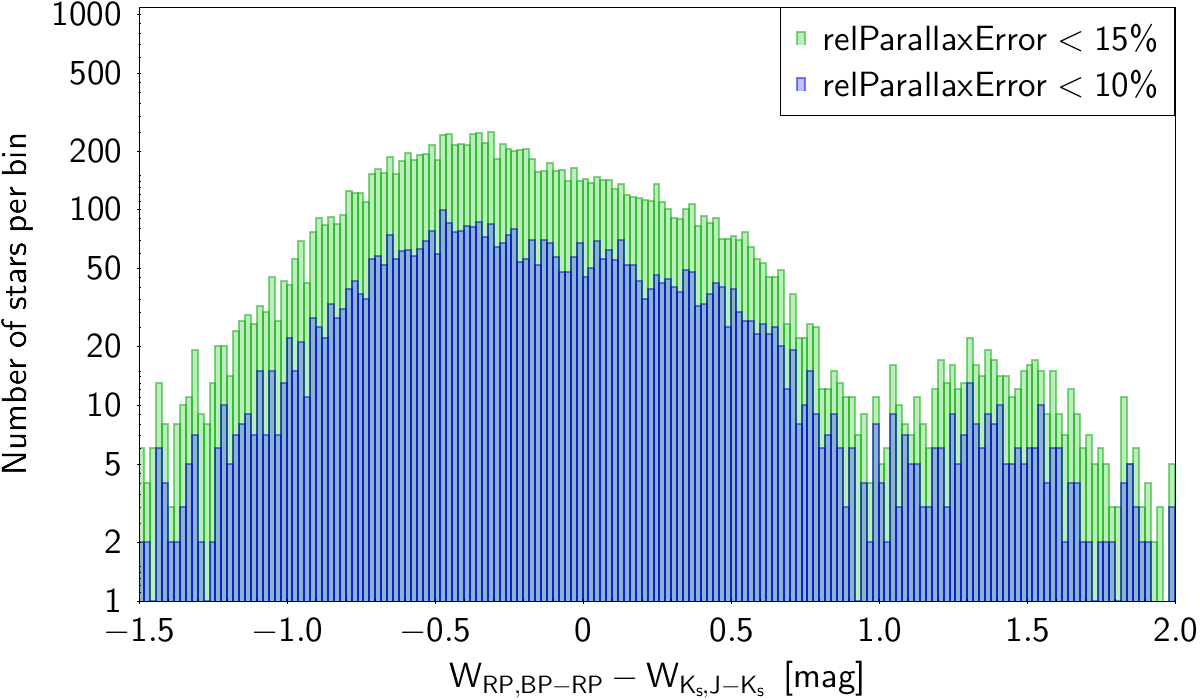}
	\caption{Histograms of the \Gaia-2MASS Wesenheit index \DeltaW of large-amplitude LPV candidates selected with $A_\mathrm{proxy}(G)\!>\!0.06$ from the full \Gaia DR2 archive. Top panel: candidates in the SMC; middle panel: in the LMC; bottom panel: in the Galaxy with relative parallax uncertainty better than 10\% in blue and better than 15\% in green.
	}
	\label{Fig:histo_WRPmWK}
\end{figure}

\begin{figure*}
	\centering
	\includegraphics[width=0.8\linewidth,angle=0]{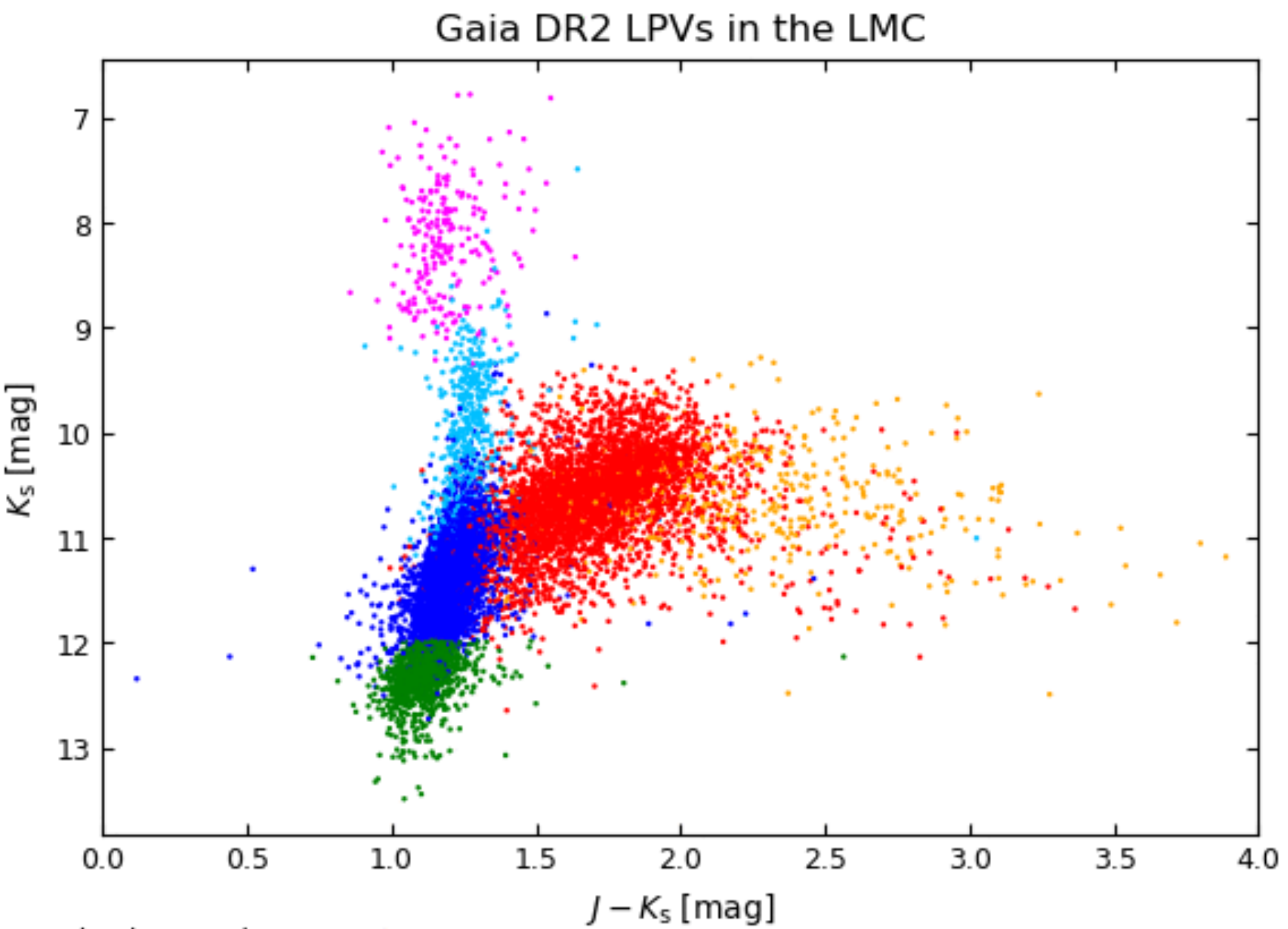}
	\vskip 8mm
	\includegraphics[width=0.8\linewidth,angle=0]{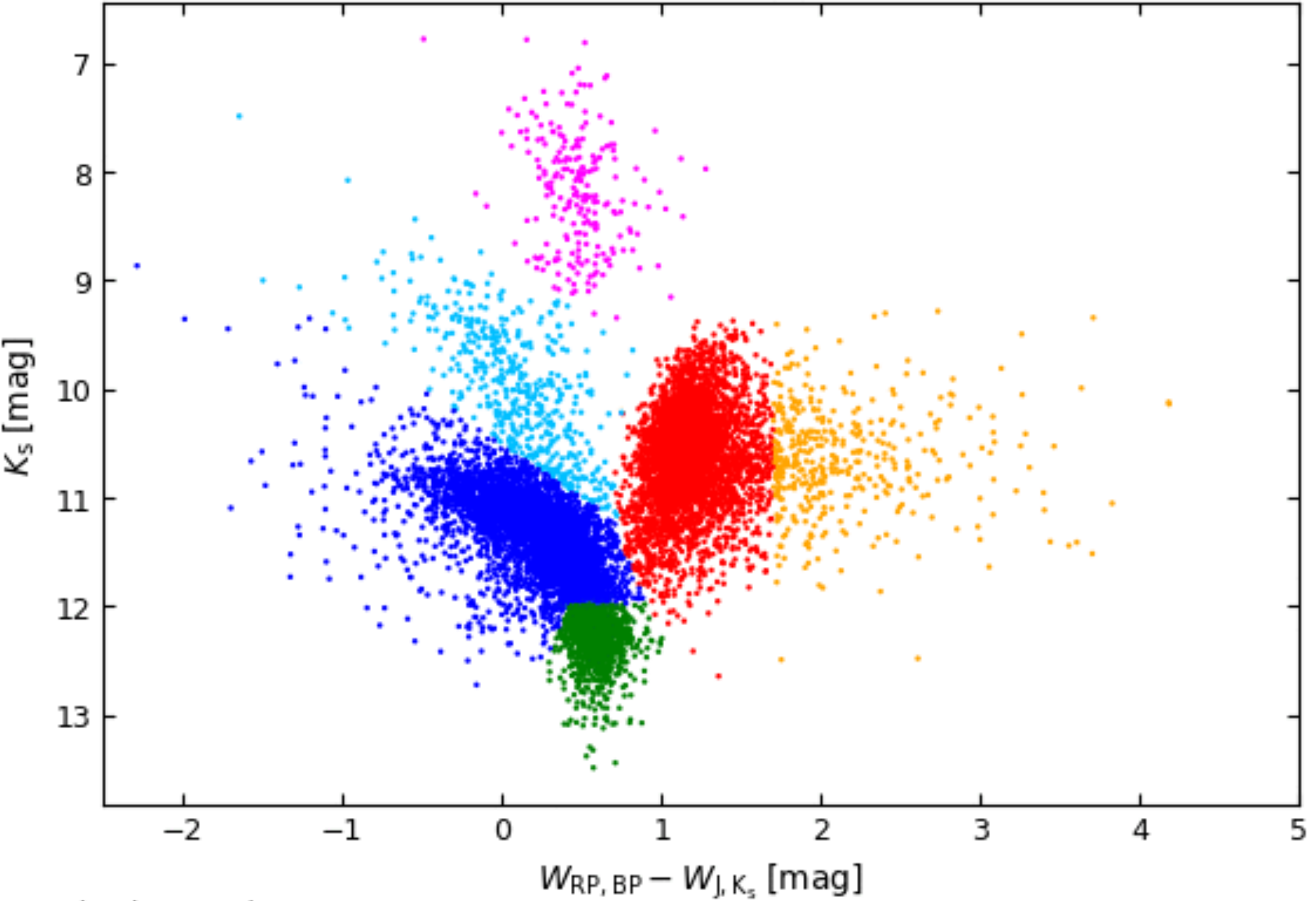}
	\caption{Distribution of the LMC LPV candidates from the \Gaia DR2 catalog of LPVs \citep{Mowlavi_etal18} in the 2MASS color-magnitude diagram (top panel, units of both axes in magnitude) compared to their distribution in the \Gaia-2MASS diagram (bottom panel).
	The colors indicate the different subgroups of AGB stars identified in \citet{Lebzelter_etal18}.\newline
	An animated image illustrating the shift from the top to bottom diagrams is available in the Zenodo archive at {\footnotesize \url{https://zenodo.org/record/3237087}}.
	%Any future change of the location on the web of this animated image will be indicated in an updated version of these proceedings on arXiv.org e-Print archive ({\footnotesize \url{https://arxiv.org/archive/astro-ph}}).
	}
	\label{Fig:G2M}
\end{figure*}

With the samples of LPV candidates presented in these proceedings for the SMC, LMC and Galaxy, we can investigate the metallicity dependence of the delineation between C-rich and O-rich LPVs in the \Gaia-2MASS diagram.
For this purpose, the histograms of \DeltaW are shown in Fig.~\ref{Fig:histo_WRPmWK} for the three samples.
The value of \DeltaW at the minimum of the histograms approximately delineates the border between O-rich (values smaller than this limit) and C-rich (values larger than this limit) stars.
The exact value actually depends on the absolute \Ks magnitude, the border not being a straight vertical line in the \Gaia-2MASS diagram, as seen in Fig.~1 of \citet{Lebzelter_etal18}.
The histograms of the different stellar samples nevertheless suggest a slight metallicity dependence of the borders separating the different subgroups of AGB stars, this border shifting to slightly larger values of \DeltaW with an increase of the (mean) metallicity of the considered stellar sample.
Further studies, outside the scope of these proceedings, should be performed on the 2-dimension distribution of sources in the \Gaia-2MASS diagram, and distinguishing various stellar populations in the Galaxy (halo, thin/thick disk, bulge), in order to reach a full understanding of the metallicity dependence of the boarders between the different regions in this diagram.

%--------------------------------------------------------------
\subsection{Using solely \Gaia data}
\label{Sect:Cstar_IoW}

With the \Gaia IoW of 15/11/2018 published on ESA's  web pages$^1$, the \Gaia Consortium revealed the potential of identifying C stars in the next \Gaia data releases\footnote{\Gaia RP spectra are not planned to be released in DR3 (see \Gaia data release scenario at {\scriptsize \url{https://www.cosmos.esa.int/web/gaia/release}}), but the \Gaia Consortium aims to provide C-rich and O-rich star identification for LPVs.} using only \Gaia data, i.e. without the necessity of infrared information.
The IoW showed that the red spectro-photometer (RP) on board of \Gaia provides low-resolution spectra that are adequate for a discrimination between O-rich and C-rich stars, from the signature of molecular bands specific to either O-rich (mainly TiO and VO) or C-rich (mainly C$_2$, CH, and CN) cool stellar atmospheres.
The reader is referred to the explanations provided therein for more information.

The IoW also showed that the large stellar surface temperature variations of Miras over their pulsation cycles can lead to significant variations in the shapes of the spectra as a function of pulsation phase.
With \Gaia, this can be effectively handled based on the multiple epoch spectra taken for each observed source.  
The IoW illustrated this power with the RP epoch spectra recorded over the DR2 period for two Miras, the O-rich Mira T~Aqr (pulsation period $P=203$~d) and the C-rich Mira RU~Vir ($P=425$~d).
%The displayed epoch spectra reveal significant variations of the shapes of the spectra as a function of pulsation phase, due to stellar surface temperature changes of these large amplitude Miras.
The links between these RP spectral variations and the photometric variability are shown in Figs.~\ref{Fig:TAqr} and \ref{Fig:RUVir} for T~Aqr and RU~Vir, respectively.
Animated versions of the figures are available on the web as mentioned in the caption of Fig.~\ref{Fig:TAqr} .

%--------------------------------------------------------------
\section{Conclusions}
\label{Sect:conclusions}

\citet{Lebzelter_etal18} have shown the power of combining  optical \Gaia and infrared 2MASS data to identify sub-groups of AGB stars among LPVs, including the distinction between C-rich and O-rich stars.
The main tool was the \Gaia-2MASS diagram shown in Fig.~\ref{Fig:LMC_Gaia2MASS_archive}.
They based their study on the LMC LPVs published in the \Gaia DR2 catalog of LPVs \citep{Mowlavi_etal18}.

In these proceedings, we first showed how the study can be extended to the set of LPVs from the whole \Gaia DR2 archive using a variability amplitude proxy based on the published uncertainties of the mean \gmag fluxes.
We then explored the \Gaia-2MASS diagram for \textit{a)} the set of LPVs in the LMC extracted from the full \Gaia DR2 archive, \textit{b)} the SMC, and \textit{c)} the Galaxy.

In the LMC, the \Gaia-2MASS diagram of the extended set of LPVs confirms the properties set forth in \citet{Lebzelter_etal18}.
In addition, it is shown (see Fig.~\ref{Fig:LMC_Gaia2MASS_archive_withVarProxy_selectedVarProxies}) that the extreme C-rich branch is almost exclusively populated by Miras, and that O-rich Miras display a spread in the diagram due to both physical properties and observational biases (see Sect.~\ref{Sect:LMC}).

In the SMC, the \Gaia-2MASS diagram displays striking differences compared to the LMC, attributed to the lower metallicity of the SMC (see Sect.~\ref{Sect:SMC}).

The study in the Galaxy is more challenging than in the Clouds due to the differential reddening affecting Galactic sources.
The analysis of the \Gaia-2MASS diagram has nevertheless led to interesting conclusions, one of them being that the clear distinction between O-rich and C-rich stars remains valid even in the presence of reddening (see Sect.~\ref{Sect:Galaxy}).

\begin{figure*}
	\centering
	\includegraphics[width=\linewidth]{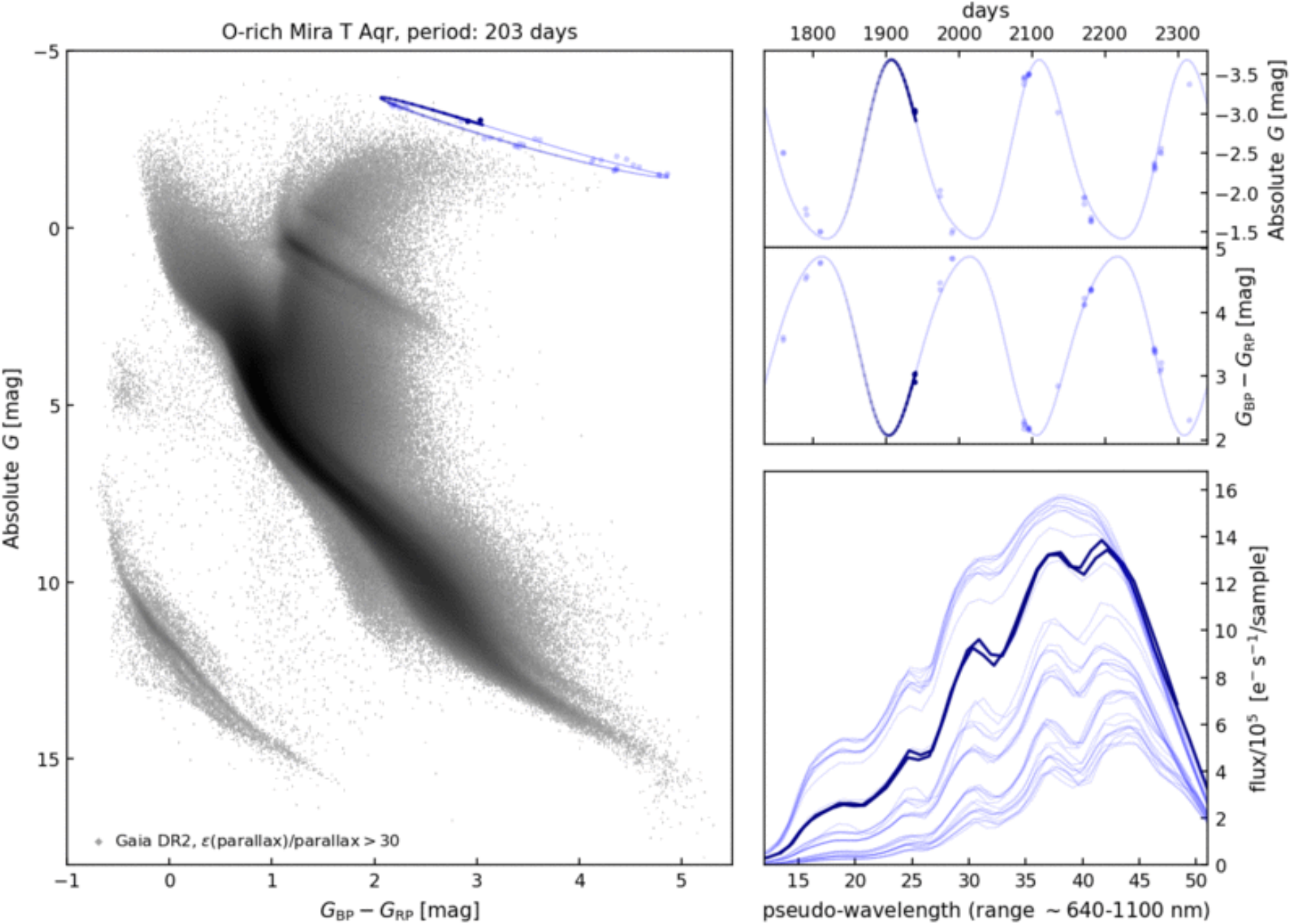}
	\caption{Photometric and spectroscopic variability of the O-rich Mira T~Aqr ($P=203$~d) during the 22 months of data covered by \Gaia DR2.
	\textbf{Left panel:} Photometric variability of the star (in blue) in the \Gaia color -- Absolute \gmag magnitude diagram.
	The background stars are the set of \Gaia DR2 sources brighter than $\gmag=18$ with parallax over parallax error ratios $\varpi/\varepsilon(\varpi)>30$ and good BP+RP excess factors.
	\textbf{Right top panel:} Time series of the \gmag magnitude and \gbp-\grp color of the star, from the \Gaia DR2 archive.
	%The times on the X-axis are given in days.
	The blue lines are Fourier series fits to the \Gaia DR2 measurements (shown as blue filled circles) in each of the two time series.
	The blue line in the left panel results from these fits.
	\textbf{Right bottom panel:} Epoch RP spectra of the star from the \Gaia IoW of 15/11/2018 (the credits are the same as in the IoW, see {\footnotesize \url{https://www.cosmos.esa.int/web/gaia/iow\_20181115}}). 
	The X-axis is a pseudo-wavelength covering the approximate range from 640 to 1100~nm.
	The dark blue points/spectra identify the same times in all panels.\newline
	An animated version of the figure is available in the Zenodo archive at {\footnotesize \url{https://zenodo.org/record/3237087}}.
	}
	\label{Fig:TAqr}
\end{figure*}

\begin{figure*}
	\centering
	\includegraphics[width=\linewidth]{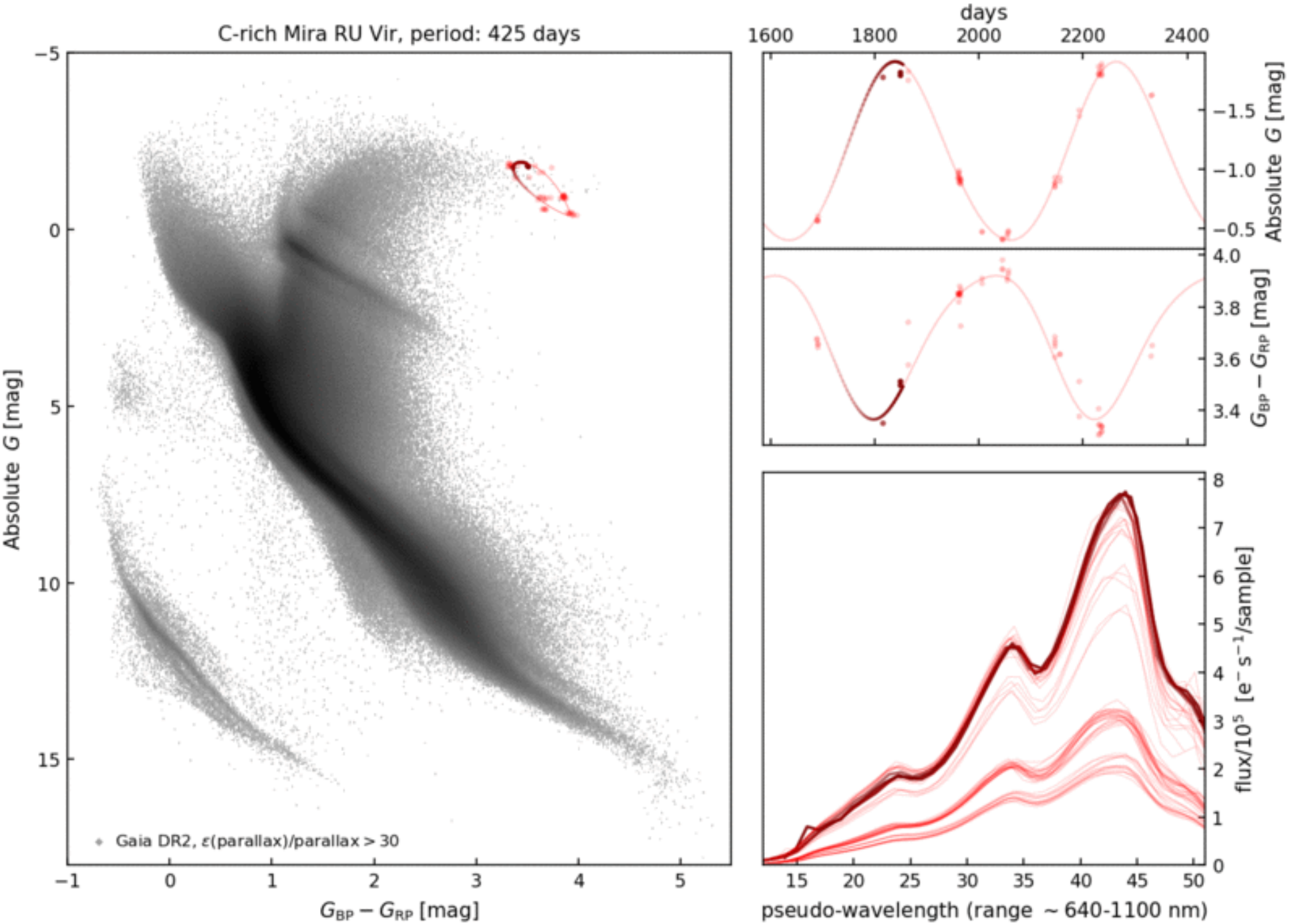}
	\caption{Same as Fig.~\ref{Fig:TAqr}, but for the C-rich Mira RU~Vir ($P=425$~d).
	The blue color of Fig.~\ref{Fig:TAqr} has been replaced by red color in all panels.
	}
	\label{Fig:RUVir}
\end{figure*}

Finally, we showed in Sect.~\ref{Sect:Cstars_diagram} that the boundary separating O-rich and C-rich LPVs in the \Gaia-2MASS diagram depends slightly on metallicity.
We then ended in Sect.~\ref{Sect:Cstar_IoW} with the interesting prospect revealed in the \Gaia IoW of 15/11/2018 that C-rich and O-rich AGB stars can be identified using \Gaia RP spectra, thereby removing the need to rely on infrared data.
This would allow to have a homogeneous set of C-rich AGB star candidates based solely on \Gaia data in forthcoming \Gaia data releases.
The IoW also showed the essential variability of the spectral features of Miras, especially of O-rich Miras.
To illustrate this, we produced animated images (Figs.~\ref{Fig:TAqr} and \ref{Fig:RUVir}) of the simultaneous photometric and RP spectroscopic variations of the two Miras shown in the IoW, that are adequate for public outreach.

%--------------------------------------------------------------
\section*{Acknowledgments}
{We would like to thank the whole \Gaia consortium for the immense work performed in the processing and analysis of the \Gaia data and their successive releases to the scientific community.
Two of us (NM and TL) are part of the \Gaia Data Processing and Analysis Consortium (DPAC), and responsible therein for the analysis and publication of LPVs within Coordination Unit 7 (CU7).
In this respect, we would like to take this opportunity to thank more specifically \textit{a)} the CU5 team in Cambridge responsible for the photometry and spectroscopy, \textit{b)} the CU7 and Data Processing Center teams in Geneva responsible for the processing of variable objects, and \textit{c)} in particular our colleague Isabelle Lecoeur-Ta\"ibi therein who actively contributes with us in the processing and analysis of LPVs.
Credits for the epoch RP spectra shown in these proceedings are the same as in the ESA/\Gaia IoW of 15/11/2018$^1$, from where they were taken, and we thank the permission to reproduce them here.
We further thank Maria S\"uveges for her help in the preparation of these spectra data for the production, by one of us (MT), of the movies and animated images accompanying Figs.~\ref{Fig:TAqr} and \ref{Fig:RUVir}.
Finally, thanks go to Walter Nowotny for additional model computations performed to investigate the effects of reddening on the properties in the \Gaia-2MASS diagram.
MT acknowledges the support from the ERC Consolidator Grant funding scheme ({\em project STARKEY}, G.A. n. 615604).\\
This work has made use of data from the European Space Agency (ESA) mission \Gaia ({\footnotesize{\url{https://www.cosmos.esa.int/gaia}}}), processed by the \Gaia Data Processing and Analysis Consortium (DPAC, {\footnotesize{\url{https://www.cosmos.esa.int/web/gaia/dpac/consortium}}}).
Funding for the DPAC has been provided by national institutions, in particular the institutions participating in the \Gaia Multilateral Agreement.
This publication makes use of data products from the Two Micron All Sky Survey, which is a joint project of the University of Massachusetts and the Infrared Processing and Analysis Center/California Institute of Technology, funded by the National Aeronautics and Space Administration and the National Science Foundation.
}

\bibliographystyle{eslab53}
\bibliography{eslab53.bib}

\end{document}